известные модели: квазисвободной частицы в «ящике» и квантового гармонического осциллятора (Ф. Блох, 1932).

*Ключевые слова:* гармонический квантовый осциллятор, квантовая частица в ящике.

## Poeschl – Teller quantum oscillator and its limiting cases

*Yuri G. Rudoy (1) (rudikar@mail.ru),*
*Enock O. Oladimedji (2) (nockjnr@gmail.com)*
*(1,2) People's Friendship University of Russia,*
*Department of Theoretical Physics and Mechanics*

*Abstract.* One of the most interesting models of the nonrelativistic quantum mechanics (in one-dimensional case) introduced by G. Poeschl and E.Teller at 1933 is considered. It is shown that this model possesses as its limiting cases two most popular quantum models: the quasi-free quantum particle in the box as well as the quantum harmonic oscillator introduced by F. Bloch at 1932.

*Keywords:* quantum harmonic oscillator, quantum particle in the box.

ВВЕДЕНИЕ. В связи с интенсивным развитием нанотехнологий (см., например, [1]), в настоящее время вновь повысился интерес к описанию квантовых объектов, состоящих из одной или нескольких микрочастиц и находящихся во внешнем силовом потенциале. Если ранее подобные задачи встречались в основном в физике атомного ядра и элементарных частиц, то сейчас они стали характерны для проблем физики конденсированного состояния, а также для объектов, находящихся в искусственных квантовых ловушках. Типичным примером могут служить *квантовые точки* в гетероструктурах [1], причем подобные квантовые объекты являются, как правило, низкоразмерными $D<3$ (ниже мы ограничимся одномерным случаем $D=1$). Для указанного класса задач представляет интерес, во-первых, наличие т.н. *управляющих параметров* – например, массы частицы $m$ и интенсивность потенциала $V_0$, и, во-вторых, возможность «термодинамической» постановки всей задачи в целом. Последнее предполагает пространственную ограниченность, или *конфайнмент*, квантового объекта в области с характерной шириной $L$, а также возможность помещения указанного квантового объекта в термостат с температурой $T$ (по шкале Кельвина). Существенно, что лишь для ограниченных объектов (с конечным $L$) имеется возможность введения понятия *механического давления* $P$, что делает «термодинамическую» постановку задачи более последовательной – в частности, позволяет проводить замкнутые термодинамические циклы.

Наиболее известными (см., например, [2-4]) – и притом точно решаемыми – задачами одномерной квантовой механики, являются следующие две задачи: 1) задача о (квази)свободной частице в «ящике» шириной $L$, а также 2) задача о гармоническом квантовом осцилляторе с собственной частотой $\omega_0$, определенная на всей оси ($L\to\infty$) – т.н. «осциллятор Блоха», введенный Ф. Блохом в 1932 году. Очевидно, что лишь первая из задач обладает конфайнментом (и, следовательно, давлением), который обеспечивается непроницаемыми для частицы стенками «ящика», что формально соответствует т.н. *дельтаобразному* потенциалу, бесконечно высокому на стенках и равному нулю внутри ящика; разумеется, подобный потенциал носит несколько искусственный характер, но допускает приближенный термодинамический анализ. Что касается второй задачи, то ее решение в «термодинамическом» варианте при ненулевой температуре $T$ является одним из немногих (если не единственным) примером *точного* квантово-статистического расчета (см., например, [5], §37). В связи с изложенным, на наш взгляд, представляет несомненный интерес рассмотрение еще одной одномерной кван-тово-механической задачи с потенциалом, предложенным Пёшлем и Теллером в 1933 году, который объединяет две указанные выше задачи и при этом допускает точное решение, описанное в двух широко известных учебных пособиях по квантовой механике [2-4]. Заметим, что с математической точки зрения задача об осцилляторе Пёшля-Теллера не намного сложнее по сравнению, например, с задачей об осцилляторе Блоха, однако с физической точки зрения указанная задача является значительно более общей и содержательной.

ПОСТАНОВКА ЗАДАЧИ. В нерелятивистской квантовой механике в одномерном случае волновая функция $\psi(x)$ и энергетический спектр частицы $E_n$ описывается стационарным уравнением Шрёдингера с потенциалом $V(x)$. Если потенциал $V(x)$ неограничен сверху и стремится к *бесконечным* значениям на концах области определения – например, на всей оси $x$ для осциллятора Блоха (БО) или на каком-либо отрезке с конечной шириной $L$ для «свободной» частицы в ящике (СЧ), то волновая функция $\psi(x)$ обращается в нуль при $x \to \pm\infty$ (для случая БО) или при $x_\pm (L) = \pm\frac{1}{2}L$ (для случая СЧ). Благодаря этому спектр $E_n$ является чисто *дискретным* и, как и потенциал $V(x)$, также неограниченным сверху, так что главное квантовое число $n$ принимает все значения от 0 (или 1) до $\infty$. Именно подобная ситуация в целом характерна для модели сильно нелинейного (по существу, даже сингулярного) осциллятора Пёшля–Теллера (ОПТ), потенциал которого имеет следующий вид, где *эффективная ширина L* имеет смысл управляющего параметра:

$$V_{\text{ПТ}}(x;L) = V_0 \text{tg}^2[\alpha(L)x], \quad 0 \leq V_0 < \infty; \quad \alpha(L) = \pi/L; \quad x_-(L) \leq x \leq x_+(L), \quad x_\pm(L) = \pm\frac{1}{2}L, \quad 0 < L < \infty. \qquad (1)$$

Потенциал $V(x;L)$ обладает рядом весьма простых и полезных свойств, которые существенно упрощают решение уравнения Шредингера для ПТ-осциллятора. Прежде всего, потенциал $V(x;L) = V(-x;L)$ является *симметричным*, или четным, относительно точки $x=0$, причем эта точка является единственным (и, следовательно, абсолютным) *минимумом*, в котором $V(0;L)=0$ для всех $L$; кроме того, потенциал $V(x;L)$ обладает парой симметричных *сингулярностей* $V(x;L) \to \infty$ для значений $x_\pm(L)$, поскольку $\text{tg}(\pi/2) \to \infty$. Наконец, потенциал $V(x;L)$ и все его производные $V^{(n)}(x;L)$ по $x$ порядка $n \geq 1$ являются *гладкими* для всех значений $x_-(L) < x < x_+(L)$ при всех отличных от нуля значениях *эффективной ширины L*, имеющей смысл управляющего параметра. Заметим, что зависимость величины $\alpha(L) = \pi/L$, $0 < \alpha(L) < \infty$, непосредственно определяющей потенциал $V(x;L)$ из (1), имеет свойство убывания при возрастании ширины $L$, так что $d\alpha(L)/dL = -\alpha(L)/L < 0$ для всех $L$, причем $\alpha_{\min}(\infty)=0$ при $L \to \infty$. Покажем теперь, что модель ПТО при определенных условиях сводится к ранее упомянутым известным моделям, а именно к модели БО вблизи точки $x=0$ для малых значений $x/L \ll 1$ и, соответственно, к модели СЧ в ящике вблизи точек $x_\pm(L)$, т.е. при малых значениях $|y|/L = |x|/L - \frac{1}{2} \ll 1$; рассмотрим эти два случая последовательно.

МЕТОДИКА. 1. Переход от осциллятора Пёшля–Теллера к осциллятору Блоха. Учтем, что при фиксированном значении $\alpha(L) = \pi/L$ и при малых значениях $\alpha(L)x \ll 1$ имеет место разложение $\text{tg}[\alpha(L)x] \approx \alpha(L)x + \frac{1}{3}[\alpha(L)x]^3 + \ldots$, так что вблизи минимума $x=0$, т.е. для значений $|x|/L \ll 1$, потенциал $V_{\text{ПТ}}(x;L)$ принимает вид:

$$V_{\text{ПТ}}(x;L) \approx V_0[\alpha(L)x]^2 \{1 + \frac{2}{3}[\alpha(L)x]^2 + \ldots\}. \qquad (2)$$

Очевидно, что потенциал (2) учитывает не только слагаемое вида $x^2$, обеспечивающее поведение типа гармонического осциллятора, но также и относительно малое слагаемое вида $x^4$, обеспечивающее поведение типа *ангармонического* осциллятора.

Потенциал (2) является интересным объектом исследования, однако здесь важно, что в (2) предполагается *конечное* значение ширины $L$ (и, соответственно, $\alpha(L)$), тогда как осциллятор Блоха определен только на всей оси $x$; последнее означает переход к пределу $L\to\infty$ и, следовательно, $\alpha(L)\to 0$.

Однако если ограничиться только этим пределом, то из (2) следует, что $V(x;L)\to 0$ при всех $x$, что не представляет интереса в термодинамическом контексте, поскольку такой предел соответствует случаю полностью свободной частицы, которая описывается волной де-Бройля и непрерывным спектром энергии. Однако ситуация изменяется, если *одновременно* с пределом $L\to\infty$ и $\alpha(L)\to 0$ перейти к пределу $V_0\to\infty$, так чтобы предел произведения $V_0\alpha^2(L)$ был *конечен*. Положив его равным $\frac{1}{2}k$=const, где $k$ – эффективный коэффициент упругости, нетрудно придти к решению для осциллятора Блоха, обладающего собственной частотой $\omega_0=(k/m)^{1/2}$.

Таким образом, предельный переход от осциллятора Пёшля–Теллера к осциллятору Блоха описывается выражением для чисто параболического потенциала на всей оси $x$:

$$V_{\text{ПТ}}(x;L) \to V_{\text{ПТ}}(x;\infty)\equiv V_{\text{БО}}(x)=\tfrac{1}{2}kx^2 \;(-\infty < x < \infty);\; k=2\pi^2(V_0/L)^2=\text{const при } V_0, L\to\infty. \quad (3)$$

2. Переход от осциллятора Пёшля–Теллера к квазисвободной частице в ящике. Как уже отмечалось выше, исходный потенциал (1) становится *сингулярным* в двух точках $x_{\pm}(L)$, что фактически означает переход к условиям обычного ящика с непроницаемыми стенками. С физической точки зрения модель Пёшля–Теллера предпочтительнее указанного «ящика», поскольку в ПТ-модели потенциал (1) сам «регулирует» поведение квантовой частицы и не нуждается в дополнительном введении весьма искусственных «стенок». Как известно (см., например, [2-4]), в модели «ящика» внутри него вообще отсутствует какой-либо потенциал, а квантование энергии частицы всецело определяется физическими *граничными условиями*, согласно которым волновая функция $\psi(x)$ строго обращается в нуль в точках $x_{\pm}(L)$. Заметим, что квантование энергии – прежде всего, строгую положительность энергии $E_1(L)$ основного состояния частицы в ящике и дискретность всего спектра $E_n(L)$

$$E_1(L)\equiv W(L)=(\hbar^2/2m)\alpha^2(L), \; E_n(L)=E_1(L)n^2 \;(n=1,2,\ldots) \quad (4)$$

можно объяснить на основе соотношения неопределенностей в форме Гейзенберга (см., например, [2-4]). Напомним, что *формально* потенциал (квази)свободной частицы в ящике может быть описан выражением

$$V_{\text{СЧ}}(x;L)=\tfrac{1}{2}\{\delta[x-x_+(L)]+\delta[x-x_-(L)]\}=\delta\{x^2-[x_\pm(L)]^2\}=\delta[x^2-\tfrac{1}{4}L^2],\; x_+(L)=-x_-(L)=\tfrac{1}{2}L. \quad (5)$$

Обратим внимание на то, что в выражение (5) вообще не входит какая-либо «амплитуда» потенциала, что означает возможность вообще положить ее равной нулю $V_0=0$; действительно, внутри ящика это так (по определению модели), а на стенках сингулярность потенциала $V_{\text{СЧ}}(x;L)$ столь сильна, что перекрывает любой предел $V_0\to 0$.

С учетом этих соображений аппроксимируем исходный потенциал ПТ-модели (1) вблизи граничных точек $x_{\pm}(L)$, перейдя к новой переменной $y_\pm=[x-x_\pm(L)]$, которая, очевидно, мала вблизи точек $x_\pm(L)$. Учтем далее, что $\alpha(L)x=\alpha(L)[y+x_\pm(L)]=\alpha(L)y\pm\pi/2$, причем $\text{ctg}(\pi/2)=0$, так что

$$\text{tg}[\alpha(L)x]=\text{tg}[\alpha(L)y\pm\pi/2]=\{\text{ctg}[\alpha(L)y]\pm[\text{ctg}(\pi/2)]\}\{\text{ctg}[\alpha(L)y]\text{ctg}(\pi/2)\pm 1\}^{-1}=\pm\text{ctg}[\alpha(L)y]. \quad (6)$$

Поскольку вблизи граничных точек величина $\alpha(L)y=\pi(y/L)\ll 1$, приближенное разложение для (6) имеет вид $\text{ctg}[\alpha(L)y]\approx[\alpha(L)y]^{-1}-\frac{1}{3}[\alpha(L)y]+\ldots$ Учитывая здесь в качестве основного только первое (расходящееся) слагаемое, получаем для потенциала (1) ПТ-модели (вблизи граничных точек) следующее приближенное выражение:

$$V_{\text{ПТ}}(x;L)\to V_{\text{ПТ}}(x\to x_{\pm}(L);L)=V_{\text{СЧ}}(y;L)\approx V_0(L/\pi)^2[1/(y_+)^2+1/(y_-)^2]=$$
$$=V_0(L/\pi)^2[(x_+(L)-x)^{-2}+(x_-(L)-x)^{-2}]=$$
$$V_0(L/\pi)^2\{(x_+(L)-x)^2+(x_+(L)+x)^2\}[(x_+(L)-x)^{-2}(x_+(L)+x)^{-2}]^{-1}=V_0(L^4/\pi^2)\{x^2-[x_{\pm}(L)]^2\}^{-2}. \quad (7)$$

Завершая достаточно протяженный (хотя и вполне элементарный) расчет (7), получаем следующий результат:

$$V_{\text{ПТ}}(x;L)\to V_{\text{ПТ}}(x\to x_{\pm}(L);L)=V_{\text{СЧ}}(x;L)\approx V_0(L^4/\pi^2)[x^2-\frac{1}{4}L^2]^{-2}; \quad (8)$$

сравнивая выражение (8) для потенциала частицы в ящике с его «точным» выражением (5), нетрудно видеть, что потенциал (8) обладает достаточно ярко выраженную сингулярность, по структуре совпадающую с (5), так что в пределе $|x|\to\frac{1}{2}L$ можно *одновременно* принять $V_0\to 0$ (сравните с аналогичным двойным пределом в выражении (3)).

3. Волновые функции и энергетический спектр квантовой модели Пёшля–Теллера. Уравнение Шредингера для одномерной модели ПТ-осциллятора имеет, как обычно, вид дифференциального уравнения 2-го порядка $(d^2/dx^2)+(2m/\hbar^2)[E-V_{\text{ПТ}}(x;L)]\psi(x)=0$ с потенциалом $V_{\text{ПТ}}(x;L)$ из (1), точное решение которого можно найти в [2-4]. Наметим основные этапы и результат этого решения. Переходя к безразмерным переменным

$$\xi=\alpha(L)x=(\pi/L)x, \ \nu(L)=V_0/W(L), \ \varepsilon(L)=(E+V_0)/W(L)=[E/W(L)]+\nu(L), \quad (9)$$

где $W(L)$ – энергия основного состояния (4) для частицы в ящике. Тогда с учетом равенства $\text{tg}^2\xi=(1/\cos^2\xi)-1$ уравнение Шредингера для волновой функции основного состояния $\psi_0(\xi)$ принимает вид

$$d^2\psi/d\xi^2+[\varepsilon(L)-\nu(L)/\cos^2\xi]\psi=0. \quad (10)$$

Для свободной частицы (при $V_0=0$, $\nu(L)=0$) в ящике с конечной шириной $L$ решением (10) является $\psi_0(\xi)=\cos\xi$. Соответственно, при $\nu(L)\ne 0$ решение (10) должно иметь вид $\psi(\xi)=(\cos\xi)^{\lambda(L)}$, где показатель $\lambda(L)>0$, поскольку согласно граничным условиям при $x=x_{\pm}(L)$ функция $\psi_0(\xi)$ должна обращаться в нуль при $\xi_{\pm}=\pm(\pi/2)$; из (10) видно, что показатель $\lambda(L)>0$ (в частности, $\lambda(L)=1$ при $\nu(L)=0$) должен удовлетворять условию

$$\lambda(L)[\lambda(L)-1]=\nu(L), \ \lambda(L)=\frac{1}{2}\{1+[1+4\nu(L)]^{\frac{1}{2}}\}, \ d\lambda(L)/dL=(1/L)\{2\nu(L)[1+4\nu(L)]^{-\frac{1}{2}}\}. \quad (11)$$

Для общего случая возбужденных состояний можно показать [2-4], что полная волновая функция $\Psi_n(\xi)$ может быть представлена в виде произведения $\Psi_n(\xi)=C_n(\lambda(L))(\cos\xi)^{\lambda(L)}G_n^{[\lambda(L)]}(\sin\xi)$, где $G_n$ – полиномы Гегенбауэра, $C_n$ – нормировочные постоянные, однако для дальнейших термодинамических вычислений существенно знание только энергетического спектра $E_n^{\text{ПТ}}(L)$, который является чисто дискретным и неограниченным сверху ($n=0, 1, 2, 3,\ldots$):

$$E_n^{\text{ПТ}}(L)=E_n^{\text{СЧ}}(L)+E_n^{\text{ГО}}(L), \ E_n^{\text{СЧ}}(L)=W(L)n^2, \ E_n^{\text{ГО}}(L)=\hbar\omega(L)(n+\frac{1}{2}), \ \hbar\omega(L)=2W(L)\lambda(L). \quad (12)$$

Учитывая определения (9) и (11), можно представить энергетический спектр (5) в более компактном виде

$$E_n^{\text{ПТ}}(L)=W(L)[n^2+2\lambda(L)n+\lambda(L)]=W(L)\{[n+\lambda(L)]^2-\lambda(L)[\lambda(L)-1]\},$$
откуда $\varepsilon_n(L)=[n+\lambda(L)]^2$ \hfill (13)

Нетрудно показать, что предельные случаи квантовой модели Пёшля–Теллера, рассмотренные в предыдущем разделе, а именно модель и модель гармонического осциллятора Блоха, имеют место не только на уровне разложения потенциала, но и на уровне энергетического спектра $E_n^{\text{ПТ}}(L)$. Действительно, модель свободной частицы в ящике соответствует *конечному* значению $L$ и отсутствию потенциала внутри ящика ($V_0\to 0$), откуда из (11) следует, что $\lambda(L)=1$, тогда как $W(L)$ вообще не меняется. При этом из (12) формально следует, что $E_n^{\text{СЧ}}(L)=W(L)n^2$ ($n=0,1,...$), что дает для энергии основного состояния (при $n=0$) неверное значение $E_0^{\text{СЧ}}(L)=0$. Однако, как показано в предыдущем разделе (см. формулу (8) и дальнейший текст), в пределе $V_0\to 0$ энергией свободной частицы в ящике является вся энергия $E_n^{\text{ПТ}}(L)$, так что указанный выше промежуточный результат должен быть поправлен за счет вклада слагаемого $E_n^{\text{ГО}}(L)$. Действительно, для рассматриваемого случая $\lambda(L)=1$ получаем из (12), что $\hbar\omega(L)=2W(L)\lambda(L)=2W(L)$, откуда сразу следует, что $E_n^{\text{ГО}}(L)=\hbar\omega(L)(n+\frac{1}{2})=2W(L)(n+\frac{1}{2})$. Тогда окончательно находим $E_n^{\text{ПТ}}(L)=W(L)(n^2+2n+1)=W(L)(n+1)^2$ ($n=0,1,...$), или в более привычном и естественном виде $E_n^{\text{СЧ}}(L)=W(L)n^2$ ($n=1,2,...$), где $W(L)$ – известный результат для основного состояния.

Модель осциллятора Блоха получается несколько более сложным путем; действительно, в пределе $L\to\infty$, $\alpha(L)\to 0$ имеем согласно (4) $W(L)=(\hbar^2/2m)\alpha^2(L)\to 0$, так что $E_n^{\text{СЧ}}(L)=W(L)n^2\to 0$, т.е вклад в спектр, *квадратичный* по главному квантовому числу $n$ и характерный для свободной частицы в ящике, полностью исчезает. При этом в полном спектре (12) остается только *линейный* по $n$ вклад $E_n^{\text{ГО}}(L)$, характерный для гармонического осциллятора, причем в блоховском пределе (т.е. при $L\to\infty$, $V_0\to\infty$, см. предыдущий раздел) частота этого осциллятора $\omega(L)$ должна перейти в блоховскую $\omega_{БО}=(k/m)^{1/2}$, где величина $k$ определена в (3). Действительно, для осциллятора Блоха согласно условиям (3) имеем $v(L)=V_0/W(L)\to\infty$, откуда согласно (11) находим $\lambda(L)\approx(\frac{1}{2})2v(L)^{1/2}=[V_0/W(L)]^{1/2}\to\infty$, а с учетом определения (3) получаем окончательно ожидаемый результат:

$$\hbar\omega(L)=2W(L)\lambda(L)\approx 2[V_0W(L)]^{1/2}=2(\hbar^2/2m)^{1/2}[V_0\alpha^2(L)]^{1/2}=2(\hbar^2/2m)^{1/2}[\frac{1}{2}k]^{1/2}=\hbar(k/m)^{1/2}\equiv\hbar\omega_{БО}. \quad (14)$$

ОБСУЖДЕНИЕ РЕЗУЛЬТАТОВ. Было бы интересно более подробно изучить энергетический спектр потенциала ангармонического осциллятора с ангармонизмом четвертого порядка (малым при больших значениях эффективной ширины $L$). Эта модель играет роль промежуточной между двумя предельными случаями свободной частицы и блоховского осциллятора, и мы оставляем соответствующий расчет заинтересованному читателю. Заметим лишь, что влияние ангармонизма существенно изменяет характер энергетического спектра, который перестает быть линейным (и, следовательно, эквидистантным); точнее, к линейному спектру $E_n^{\text{ГО}}(L)=\hbar\omega(L)(n+\frac{1}{2})$ в низшем порядке теории возмущений добавляется слагаемое $\Delta E_n(L)=W(L)(n^2+n+\frac{1}{2})$ (см., например, [2-4]). Правда, в более высоких порядках теории возмущений добавляются «нефизические» слагаемые порядка $n^3$, $n^4$,..., отсутствующие в *точном* выражении (12), что явно свидетельствует об ограниченной применимости теории возмущений применительно к сингулярным потенциалам, к которым относится потенциал модели Пёшля–Теллера.

ЗАКЛЮЧЕНИЕ. Содержание данной статьи можно кратко резюмировать следующим образом. Как видно из выражений (3) и (8), модель нелинейного сингулярного квантового осциллятора Пёшля–Теллера действительно объединяет две основных модели квантовой механики (в одном измерении), а именно: при низких энергиях (вблизи центра «эффективного ящика») модель ПТ-осциллятора воспроизводит модель линейного гармонического осциллятора Блоха, тогда как при высоких энергиях (вблизи стенок «эффективного ящика») модель ПТ-осциллятора воспроизводит модель (квази)свободной частицы в ящике. Эти выводы подтверждаются точным решением уравнения Шредингера

*Список использованной литературы*

УДК ????

## Обобщенный закон Дарси в теории фильтрации: пристеночный эффект

*Рыбаков Ю.П. (1) (soliton4@mail.ru),*
*Семенова Н.В. (1) (dobroe_slovo@inbox.ru)*
*(1) Российский университет дружбы народов,*
*факультет физико-математических и естественных наук*

*Аннотация.* Рассматривается течение жидкости в пористой среде для случая аксиальной симметрии. Оценивается эффективность фильтрации для фильтров двух возможных геометрий: цилиндрической и радиальной. Предлагается обобщение закона фильтрации Дарси и объясняется так называемый пристеночный эффект.

*Ключевые слова:* закон Дарси, фильтрация, поперечная диффузия, пористые среды.

## Generalized Darcy's law in filtration theory: the "near-wall" effect

*Rybakov Yu.P. (1) (soliton4@mail.ru),*
*Semenova N.V. (1) (dobroe_slovo@inbox.ru)*
*(1) Peoples' Friendship University of Russia, Faculty of Science*

*Abstract.* The liquid flow in a porous medium is considered for the axially-symmetric case. The filtration efficiency is estimated for filters of two possible geometries: cylindrical and radial ones. The generalization of the Darcy's filtration law is suggested and the explanation of the so-called "near-wall" effect is given.

*Keywords:* Darcy's law, filtration, transverse diffusion, porous media.